\documentclass[12pt]{iopart}
\usepackage{graphics}
\usepackage{epsfig}
\newcommand{\eeq}{\end{equation}}      \newcommand{\beql}[1]{\begin{equation}\label{#1}}
\newcommand{\req}[1]{Eq.\,(\ref{#1})}  
\begin{document}
\title{Strangeness enhancement at LHC}
\author{Johann Rafelski$^1$, and Jean Letessier$^{1,2}$}%
\address{$^1$ Department of Physics, University of Arizona, Tucson, AZ 85721\\
$^2$ %Laboratoire de Physique Th\'eorique et Hautes Energies\\
LPTHE, Universit\'e Paris 7, 2 place Jussieu, F--75251 Cedex 05}

%\ead{rafelski@physics.arizona.edu}  %restore if fits 6 pages <<<<<<<<<

\begin{abstract}
{We study production of strangeness in the hot QGP fireball in conditions achieved   at LHC, and use these results 
to obtain soft (strange) hadron multiplicities. We compare the chemical equilibrium and non-equilibrium conditions
and identify characteristic experimental observables.
} 
\end{abstract}
%\maketitle
%%%%%%%%%%%%%%%%%%%%%%%%%%%%%%%%%%%%%%%%%%%%%%%%%%%%%%%
\section{Introduction} \label{intror}
We address here the bulk matter hadronization at LHC-ion. 
This work represents a synthesis of insights we 
published in past 2 years~\cite{Rafelski:2005jc,Letessier:2006wn}, 
and  it comprises a comparative study of LHC bulk hadronization. One of the 
points of interest is how, as function of increasing energy of colliding nuclei, 
the hadronization of the dense  quark--gluon matter fireball  occurs --- experimental
evidence suggests that  at sufficiently high energy, e.g., at RHIC,  this happens 
 in a rapid and explosive manner. The LHC-ion research program should  
allow exploration of   even more extreme conditions.

Since QGP phase is entropy rich, compared to the hadron phase, additional entropy production
at hadronization hinders hadronization. Thus, in the explosive QGP breakup process, we expect
entropy to be preserved.   Moreover,  also the flow  of QGP  matter seems to 
occur without a significant entropy production~\cite{Muller:2007rs}. Thus, 
the entropy  (per unit rapidity) $dS_0/dy$ produced in the early reaction state is also
nearly the entropy present in the QGP breakup. This means that
the total final state hadron multiplicity prior to hadron resonance decay 
is a  measure of initial  entropy   production, which occurs predominantly   
during  the  thermalization stage~\cite{Bjorken:1982qr}. 
We will constrain the initial state conditions by considering the final state entropy
(hadron multiplicity) expected at LHC. 

Our study of  the   strange quark production processes 
is based on kinetic theory of particle collisions, and as is well 
known, the gluon collisions are driving strangeness to chemical
equilibrium in hot QGP. This seems to depend on the degree of chemical 
equilibration in the early QGP phase. However,   the observable final specific
`strangeness pair yield per entropy' $N_s/S$ 
(also colloquially referred to as $s/S$)
ratio is found to be rather  insensitive to this uncertainty. This  expresses 
the fact that, at a given entropy content,
the temperature can be high at low particle yield, or vice-versa, 
temperature can be low, but particle number  large.  In the cumulative strangeness  
production process, these effects compensate,
the final   strangeness yield is not initial state dependent (i.e., not dependent on
full gluon equilibration)~\cite{Letessier:2006wn}. This observation illustrates the fact
that strangeness is not a `deep' observable of QGP, that is we cannot use it to 
study conditions prevailing at the early stage of heavy ion reaction. 

After a brief discussion of the kinetic strangeness production which allows us to 
evaluate the expected strangeness yield,  we turn our attention to the  properties of 
the fireball of matter and its breakup (hadronization).

%%%%%%%%%%%%%%%%%%%%%%%%%%%%%%%
\section{Strangeness chemical equilibration in QGP at LHC}\label{sLHC}
%%%%%%%%%%%%%%%%%%%%%%%%%%%%%%%
Both strangeness and entropy are nearly conserved near, and at
hadronization,
and thus the final state $s/S$, is closely
related to the thermal processes  in the fireball at $\tau\simeq 1$--4 fm/c.
The measurement of $s/S$ involves study of strange and non-strange hadrons
and extrapolation of unobserved yields, required to evaluate the entropy
locked in hadron multiplicity. In order to evaluate the magnitude of $s/S$ in the QGP phase,
we consider the hot  early  stage of the reaction. For  an   equilibrated 
QGP phase with perturbative properties, we have  (superscript `G' indicates that this 
quantity is referring to QGP phase).
\beql{sdivS}
{s \over S}\equiv\frac{\rho_{\rm s}}{S/V}   \simeq
\frac{ (\gamma_s^{\rm G}(t) g_s/\pi^2) T^3 0.5\,x^2K_2(x)}
  {g\,4\pi^2/ 90\, T^3}
%    +n_{\rm f}[(7\pi^2/ 15) T^3 + \mu_q^2T]}
=\frac{\gamma_s^{\rm G} g_s}{g} 0.23 [0.5 x^2K_2(x)]\,.
%\simeq 0.046 
\eeq
For early times, when $x=m_s/T(t)$ is relatively small, the equilibrium value of strangeness QGP phase space 
occupancy ($\gamma_s^{\rm G}=1$)
can be as large as $s/S> 0.04$. However, at high temperature, strangeness is not yet equilibrated
chemically.  For $m_s/T\simeq 0.7$, appropriate for hadronization stage, 
the QGP chemical equilibrium is reached
 when  $s/S\simeq 0.04$. 

The temporal evolution of $s/S$, in an expanding plasma, 
is governed by:  
\begin{eqnarray}\label{qprod3a}
 {d\over { d\tau}} {N_s\over S}
&=&
 {A^{gg\to s\bar s}\over (S/V) } %{4\pi^2\over 90} g(T)T^3}
    \left[(\gamma_{\rm g}^{\rm G}(\tau))^2\!-(\gamma_{s}^{\rm G}(\tau))^2\right] 
% \nonumber\\ 
%&&+
+ 
{A^{q\bar q\to s\bar s}\over (S/V) } %{4\pi^2\over 90} g(T)T^3}
    \left[(\gamma_q^{\rm G}(\tau))^2\!-(\gamma_{ s}^{\rm G}(\tau))^2\right]\,.
\end{eqnarray}%{equation}
When all $\gamma_i\to 1$, the Boltzmann collision term vanishes, and
 equilibrium has been reached.  Here, we use the invariant rate per unit time and volume, 
$A^{12\to 34}$, by incorporating the  chemical equilibrium densities into the thermally
averaged cross sections:
\begin{equation}
A^{12\to 34}\equiv\frac1{1+\delta_{1,2}}
\gamma_1\gamma_2  \rho_1^\infty\rho_2^\infty 
             \langle \sigma_{s} v_{12} \rangle_T^{12\to 34}.  
\end{equation}
$\delta_{1,2}=1$ for   reacting particles being identical bosons, $\delta_{1,2}=0$ otherwise . 
$\gamma_i$ expresses the deviation
from equilibrium of density $\rho_i$. Note also that
the evolution for $s$ and $\bar s$, in proper time of the 
co-moving volume element, is identical as both change in pairs.

We evaluate $A^{gg\to s\bar s}$ and $A^{q\bar q\to s\bar s}$ employing the available
strength of the QCD coupling, and  $m_s(\mu=2\,{\rm GeV})=0.10$ GeV for the strange 
quark mass. 
We compute rate of reactions employing a running strange quark mass working
in two loops, and using as the energy scale the CM-reaction energy 
$\mu\simeq  \sqrt{s}$. For the   QCD couping constant, we
 use  as reference value    $\alpha_s(\mu=m_{Z^0})=0.118$, 
and evolve the value to applicable energy domain $\mu$ by using two loops. 
We introduce a multiplicative factor $K=1.7$,  which accounts for processes 
odd in power of $\alpha_s$, 
such as gluon fusion into strangeness, with  gluon 
bremsstrahlung emitted by one of the strange quarks.   Being distinguishable, `even' and `odd'  terms
are contributing  incoherently, always increasing the production rate. 
The magnitude of $K$, is 
estimated based on   perturbative QCD  Drell--Yan  lepton pair production~\cite{Hamberg:1990np},
and   heavy quark production~\cite{Kidonakis:2004qe}.

In order to evaluate strangeness production, we need to understand $T(\tau)$. Hydrodynamic
expansion with Bjorken scaling~\cite{Bjorken:1982qr}, and small viscosity implies 
that   $dS/dy\equiv \sigma(T) dV/dy ={\rm Const.}$ as function of time. This means that 
 $dV/dy (\tau)$ expansion fixes $\sigma(\tau)$, thus based on equations of state, also 
$T(\tau)$. We parametrize  $dV/dy\propto A_\bot dz/dy|_{\tau,y}$ where in Bjorken scaling
limit $dz/dy|_{\tau,y}=\tau \cosh y$. Thus, the ad-hoc model element is the transverse 
flow, which defines the size of the transverse fireball slice, $ A_\bot(\tau)$. Our model
explorations of `reasonable' forms of $ A_\bot(\tau)$ suggest that the controlling factor in 
the determination of $s/S$ at freeze-out is the final size of the fireball along with final velocity, 
and  not the system history. This is consistent with our earlier observation that strangeness 
is not a `deep' observable of QGP. 

The value of  $s/S$ is mainly uncertain due to unknown initial entropy $S_0$, and 
the strange quark mass. The latter uncertainty does not impact the degree of strangeness chemical
equilibration in QGP: for smaller mass, more strangeness can be made, but also more strangeness
is needed to equilibrate the higher phase space density. These two effects cancel 
nearly exactly in a wide range of model parameters. Thus, the achievement of $\gamma_s^{\rm G}\to 1$
chemical equilibrium in QGP is not a model dependent outcome,  and is in essence a result of entropy content
provided by heavy ion collisions.  However, the actual final specific yield $s/S$ varies
according to \req{sdivS}. In what follows, we assume chemical equilibration in QGP.

%%%%%%%%%%%%%%%%%%%%%%%%%%%%%%%
\section{Soft Hadrons at LHC}\label{SHLHC}
%%%%%%%%%%%%%%%%%%%%%%%%%%%%%%%
The phase space density is, in general, different in any 
two phases. Hence, when transformation of one phase into the other occurs rapidly, 
given chemical equilibrium in the decaying phase, the final state is out-of chemical 
equilibrium. Thus,  in order to preserve entropy, there must be  a jump in the phase
 space occupancy parameters $\gamma_i^{\rm G}<\gamma_i$  (When
we omit superscript on $\gamma$ this quantity refers to hadron phase space.)
This jump replaces the increase in volume found 
in a slow transformation involving  yield  re-equilibration, which is required in 
a putative chemical equilibrium approach. 
In order to preserve entropy in sudden hadronization of supercooled QGP  at  $T\simeq 140$ MeV,
we  must have  $\gamma_q^{\rm cr}=e^{m_\pi/T}$.
The value $\gamma_q^{\rm cr}$  is where the pion gas  condenses. The required value of  $\gamma_q$ is
decreasing with increasing temperature
and crossing  $\gamma_q=1$  near $T\simeq 180$ MeV.
We study the soft hadron production at LHC subject to the  following  constraints and inputs,  using 
the statistical hadronization model (SHM) implemented 
in the package SHARE~2~\cite{Torrieri:2004zz}: \begin{enumerate}
\item The  extrapolation as function of $\ln\sqrt{s_{\rm NN}}$ implies an increase
of $dS/dy$ by  a factor 1.65 at  LHC-5520 compared to RHIC-200 (numerals refer to $\sqrt{s_{\rm NN}}$).
However, we also consider the case of 3.4-fold increase in particle multiplicity, that is within a TPC the charged
track number is $h=2924$. This yield occurs in hadron gas in  chemical
equilibrium at the final volume $dV/dy=6200\,{\rm fm}^3$.
\item We take   strangeness content from  pQCD considerations discussed above, which at LHC
implies  QGP  near chemical equilibrium conditions. This can be   compared  to 
expectations assuming hadron equilibrium condition.
\item  We fix the value of $\mu_{\rm B}$ and by strangeness balance $\langle s\rangle = \langle \bar s\rangle$ 
also of $\mu_{\rm S}$. We do this   aiming
to obtain for the specific energy  per nett baryon $E/(b-\bar b)\simeq 415\pm 20$\,GeV,  a value to which 
our study of SPS and RHIC extrapolates to. We make sure that the resulting charge per nett baryon ratio
$Q/(b-\bar b)=0.4$ is as provided by the proton to neutron ratio in heavy ions.
\item We fix as hadronization condition for sudden break up model $\gamma_q=1.6$ and  $T_f=140$\,MeV;
the value of $\gamma_s$ follows from specific strangeness conservation $s/S$ in hadronization.
For the simple hadron model, we take $T_f=162$\,MeV and $\gamma_q=\gamma_s=1$.
\item We bias  our fit  towards $\pi^+/\pi^-\simeq 1$. For the non-equilibrium hadronization,
we also bias $E/TS\simeq 1$. This constrains the search of the minimum considerably and 
renders the result more numerically stable.
\end{enumerate}

In table \ref{Model}, we present  in the column on left the  properties of the fireball consistent with 
hadron yields   extrapolated using  PHOBOS-RHIC  
 charged particles yield, which at LHC yields
 $dh_{\rm charged}/dy=1150$~\cite{Busza:2007ke}.  In the second column, we
fix   the visible hadron multiplicity in STAR like set-up, to correspond to a reference value
obtained in the chemical equilibrium model, with volume fixed~\cite{Andronic:2005yp} in order to increase 
the hadron yield by about factor 3.4. This results in  enhanced entropy content, $dS/dy$ and thus, we use
somewhat higher specific strangeness content $s/S$, in accordance with our kinetic strangeness
production computation. 

The high specific strangeness yield of the chemical non-equilibrium model is expressed
by the large value $\gamma_s>1$. The low temperature at freeze-out ($T_f=140<161$\,MeV)
leads to $\gamma_q\simeq 1.6$. While this value is result of a constraint to stop condensation
of pions, one can show  that this
value  preserves entropy contents of QGP in the HG phase~\cite{Kuznetsova:2006bh}.  Despite 
the lower value of $T_f$ in the non-equilibrium model, the fireball has 
greater energy and entropy density. As the  entries in table \ref{Model}  show, 
this is in essence due to a significantly greater strange hadron contents.
Generally,  there are several distinctive hadron multiplicity features of
hadronization models. In particular, enhanced production of multistrange hadrons
occurring at the same time as somewhat suppressed resonance production,  is a strong indication
that rapid, non-equilibrium breakup of QGP  is present.

\begin{table}[!ht]
\caption{\small Fireball  properties at freeze-out, followed by particle yields (prior/after weak decays)
and ratios.  First two columns, the chemical  non-equilibrium model,  equilibrium model follows in
the last column. See text for details. 
  Star `*' indicates a fixed value, i.e. input used to characterize the model. We 
bold-face  40\%+difference of chemical non-equilibrium to equilibrium models.\\[-0.3cm]
\label{Model}}
%\footnotesize
\rm\centering
\begin{tabular}{|c| c | c || c |  }
\hline
$T$[MeV]        & $140^*$& $140^*$  & $161^*$   \\
$dV/dy$[ fm$^3$]        &2126  &4223  &6200$^*$    \\
$dS/dy$          &7457  & 16278  &  18790   \\
$ b-\bar b $  &2.6&5.5&6.4 \\
$dh_{\rm ch}/dy$ (PHOBOS) &{\bf $1150^*$}  &$2435 $      &$2538 $     \\
$dh_{\rm ch}^{\rm vis}/dy$ (STAR)&$1350\phantom{^*} $  & $2924^* \to$     &$ 2924$  \\
$(b+\bar b)/h^-$  &0.334&0.353&0.370 \\
\hline
$1000\cdot (\lambda_{\rm q,\, s}-1)$     & $5.6^*, \,2.1^* $   &$ 5.6^*, \,2.1^*$    &$ 5.6^*, \,2.0^*$     \\
$\mu_{\rm B,\, S}$[MeV]     & $2.3^*, \,0.5^*$   &$2.3^*,\, 0.5^*$    &$2.7^*,\, 0.6^*$     \\
$ \gamma_{q,\,s}$      & $1.6^*,\, 2.35$  &$ 1.6^*, \, 2.8$   &$1^*,\,1^* $   \\
%$\gamma_s $      &2.35  &2.78  &1*    \\
\hline
$s/S$              &  $0.034^*$ &$0.038^*$   &$0.0255$    \\
$E/(b-\bar b)$              &$423$ &431 &404  \\
$E/TS$              &1.04 & 1.04& 0.86 \\
$P/E$              & 0.165& 0.162&0.162  \\
$E/V$[MeV/fm$^3$ ]            &509 &560 &420\\
$S/V$[1/fm$^3$ ]            &3.51 &3.86 &3.03\\
$(s+\bar s)/V$[1/fm$^3$ ]            & 0.119 &  0.147 &0.077 \\
%$P$[MeV]   &84& 91&68\\
\hline
$0.1\cdot\pi^\pm$            &49/61  &102/132  &115/132    \\
$p$   &$25/45 $  &$50/101$    &$71/111 $     \\
$\Lambda$          &19/27 &45/70    &40/53         \\
${\rm K}^\pm$        &94   &226   &183      \\
$\phi $            &14  & {\bf  38 }   &25       \\                     %<<<<<<<
$\Xi^-$             &3.9  & {\bf 11}   &6.2   \\                 %<<<<<<<<<<<<<<<<<<<<
$\Omega^-$         &0.78 & {\bf  2.6}   &0.98     \\       %<<<<<<<<<<<<<<<<<<<
\hline
%1000$\phi/h_{\rm ch}^{\rm vis} $            &10  &\vio 13    &8.4\bla       \\                    %<<<<<<<
%1000$\Xi^-/h_{\rm ch}^{\rm vis}$             & 2.9&\vio 3.8  &2.1 \bla  \\                  %<<<<<<<<<<<<<<<<<<<<
%1000$\Omega^-/h_{\rm ch}^{\rm vis}$         & 0.58&\vio 0.88    &0.36 \bla    \\    %<<<<<<<<<<<<<<<<<<<
\hline
$\Delta^{0},\,\Delta^{++} $      
                 &4.7   &{\bf  9.4}   &14.6        \\
$K^*_0(892)$ & 22&52&55 \\
$\eta$ &62&149&133 \\
$\eta'$ & 5.2 &13.2&12.1\\
$\rho$ &36& {\bf  74} &119  \\
$\omega$ &32& {\bf   65}& 109   \\
$f_0$ & 2.8& {\bf  5.6 }&10.2  \\
\hline
K$^+ /\pi^+_{\rm vis}$   & 0.164 &0.184  &  0.148 \\
$\Xi^-/\Lambda_{\rm vis}$ &0.143 & {\bf 0.159}& 0.116\\
$\Lambda(1520)/\Lambda_{\rm vis}$ &0.044 &{\bf  0.041} &0.060   \\
$\Xi(1530)^0/\Xi^-$ &0.33 &0.33 &0.36 \\
1000$\phi/h_{\rm ch}^{\rm vis} $            &10  &{\bf  13}    &8.4        \\                    %<<<<<<<
%$\phi/{\rm K}^+$ &0.15 & 0.17&0.135 \\
$K^*_0(892)/K^-$ & 0.237&{\bf 0.232} &0.303 \\
\hline%\end{tabular}%\vspace*{0.1cm}
\end{tabular}
\end{table}

%%%%%%%%%%%%%%%%%%%%%%%%%%%%%%%%%%%%%%%%%%%%%%%%%%%%%
\section{Discussion} 
%%%%%%%%%%%%%%%%%%%%%%%%%%%%%%%%%%%%%%%%%%%%%%%%%%%%%
 One is tempted to wonder if and why the use of chemical non-equilibrium parameters matters. 
Note that  the yield of  baryons is proportional to $V \gamma_q^3 (\gamma_s/\gamma_q)^{n_s}$,
where $n_s$ is the total number of constituent  strange quarks or anti-quarks. Similarly,  
the yield of mesons is proportional to $V \gamma_q^2 (\gamma_s/\gamma_q)^{n_s}$.
Thus, ratio of baryons to mesons at   fixed value of $\gamma_s/\gamma_q$
is  dependent on the value of $ \gamma_q$. The parameter
$ \gamma_q$ allows us to increase the yield of baryons compared to the yield
of mesons. This is a very important feature  considering
that microscopic pictures of quark recombination.

Given the mass difference between mesons and baryons,  $T_f$ also
impacts the relative meson to baryon ratio. However,  
solely $T_f$  influences the relative resonance   yield, and should for this reason not
be ad-hoc  manipulated to balance the baryon to meson yield.  
A consequence of fixing (arbitrarily) $\gamma_q=1$ is that the   overall 
ratio of baryons to mesons fixes the value of $T_f$ and predicts 
resonance yields which, in general, at RHIC disagree with experiment
by up to 50\%. This (in part) explains why,  when $\gamma_q\ne 1$, one finds fits 
of experimental data which are a lot better than for $\gamma_q=1$. We note that
there have been efforts to explain the reduced yields by  rescattering of daughter particles. 
Our sudden hadronization approach is simpler and the outcome has no fine-tuned reaction parameters.
It is possible  using double resonance ratios to  differentiate these models.
As a third point,  ad-hock assumption of a value expected in a perfect world, viz.   
chemical hadron equilibrium  $ \gamma_q=1$ 
presumes  also that the SHM model is completely understood. In our opinion
$ \gamma_q$  a) helps to compensate 
 model omissions (e.g., hadron widths not considered in general, though available in SHARE), 
and, b)  corrects for evolving but at present still incomplete knowledge of hadron spectra.
We conclude that introduction of $\gamma_q$, within the SHM, is a necessary model feature.

Another LHC soft hadron abundance controlling parameter is $\gamma_s/\gamma_q$, which 
influences all strange to non-strange particle yields, 
allows to preserve strangeness yield across the phase boundary. At LHC, we expect greater
intrinsic strangeness content, and thus, a relatively large  value of $\gamma_s/\gamma_q$,
which enters with power $n_s$ for the number of valance strange quark we compare non-strange
hadrons with. Since strange hadrons are in general more massive than non-strange, the higher 
equilibrium model $T_f\simeq 161$ MeV in part compensates the model strangeness suppression
for singly strange hadrons, but multistrange hadrons allow to discriminate models, and thus, to 
understand the properties of QGP at hadronization, and hadronization dynamics.

\subsubsection*{Acknowledgments}
Work supported by a grant from: the U.S. Department of Energy  DE-FG02-04ER4131.
LPTHE, Univ.\,Paris 6 et 7 is: Unit\'e mixte de Recherche du CNRS, UMR7589
%%%%%%%%%%%%%%%%%%%%%%%%%%%%%%%%%%%%%%%%%%%%%%%%%%%%%

\end{document}